\documentclass[fleqn,12pt]{article}

\begin{document}
\vspace*{-0.3cm}
\begin{center}
{\large\bf Some examples of uses of Dirac equation and its
generalizations in particle physics }

\bigskip
\medskip

{\large V.V. Khruschov}

\medskip

{\it
  RRC "Kurchatov Institute", Ac. Kurchatov Sq. 1,
 Moscow, 123182, Russia
}

\vspace*{-0.5cm}

\begin{abstract}
Applications  of  the  Dirac equation with an  anomalous  magnetic
moment are  considered for description of characteristics of electrons, 
muons  and quarks. The Dirac equation with  four-dimensional
scalar and vector potentials is reduced to a form suitable  for  a
numerical  integration. When a certain type of  the  potential  is
chosen,   solutions can approximate quark  states  inside
hadrons. In view of complicated behaviour of quarks  in
a  confinement domain some generalizations are considered such  as
the  Dirac-Gursey-Lee  equation, the Dirac  equation  in  a  
five-dimensional Minkowski space, the Dirac equation in a quantum phase
space.  Extended  symmetries  for  the  Dirac  equation  and   its
generalizations are considered, which can be used for  investigation
of  properties  of  solutions of these  equations  and  subsequent
applications in particle physics.

\end{abstract}
\end{center}

\bigskip
\medskip

\noindent {\bf 1. Introduction}

   It  is  well  known   the  Dirac equation  for  a  relativistic
electron  is  one of the main building blocks of the quantum electrodynamics
(QED)  and the Standard Model (SM) [1, 2]. It is used to  check
the  validity  of  SM  as  to search for new  physics  beyond  SM.
Although  the  Dirac  equation was offered to the  description  of
electron  it  describes with success muons,  tayons  and  moreover
quarks.  The  fact  of  validity of the Dirac  equation  (DE)  for
description of quarks is astonishing because quarks do not observe
in our physical spacetime.
   DE   is  used  in  the  quantum  chromodynamics  (QCD) [2]  and  in
phenomenological  hadron models, for example, in the  relativistic
model  for  quasi-independent quarks (RMQIQ), which has  been
applied for the description of hadron properties [3,  4]. However
in  a  domain  of confinement the deviations  from
the Dirac equation and standard spacetime symmetry can take place.
   From  this point of view generalizations of DE are welcome.
We  consider  some generalized equations of Dirac type,
namely,  the Dirac-Gursey-Lee equation (DGLE), the Dirac  equation
in  a  five-dimensional Minkowski space (FDDE), the Dirac equation
in a quantum phase space (QPSDE).

\smallskip

\noindent{\bf 2. The Dirac equation and  AMMs of electrons, 
muons and quarks}

     In  1928  Dirac presented the electron equation, which is
used successfully for description of fermions at present. Let us
write DE in the natural system of units: $c = \hbar = 1$ 
($p^i =i\partial^i$, $ x^i =\{t,\mathbf{x}\}$,
$\gamma^i=\{\beta,\beta\mathbf{\alpha}\}$).
\begin{equation}
  (\gamma^ip_i - m)\psi(x) = 0,
   \quad\beta =
\left(\begin{array}{cc}\sigma_0 & 0\\
0 & -\sigma_0\end{array}\right), \quad \alpha =
\left(\begin{array}{cc}0&\sigma\\
 \sigma&0\end{array}\right),
\label{newd}
\end{equation}
\noindent where $\sigma_0 =1_2$, $\sigma$ are the known Pauli matrices, the
metrical tensor  $\eta_{ij} =$  $ diag \{1, -1,$ $ -1, -1\}$.
     DE  is  invariant  under transformations of  Poincare  group,
which  are generated  by  operators of momenta $p^i$ and  four-dimensional
rotations
\begin{equation}
                 J_{ij} = x_i p_j - x_j p_i  + S_{ij},  \quad
 S_{ij} = \frac{i}{2}\sigma_{ij}, \quad
   \sigma_{ij}  = \frac{1}{2}(\gamma_i\gamma_j  - \gamma_j\gamma_i).
\end{equation}

If we define the pseudoscalar matrix $\gamma_5 =
i\gamma^0 \gamma^1 \gamma^2 \gamma^3$,  the left and
right  wave functions $\psi_L$, $\psi_R$  can be written
\begin{equation}
   \psi_L =  \frac{1-\gamma_5}{2}\psi_,
\quad\psi_R  =  \frac{1+\gamma_5}{2}\psi_,
\quad\gamma_5\psi_{R,L}  = \pm\psi_{R,L}.
\end{equation}
One can consider the charge conjugated fields  $\psi^c=C(\bar\psi)^T$,
where $C$ is
the matrix of the charge conjugation which satisfies the relations:
$  C\gamma_i^TC^{-1} = -\gamma_i$, $ C^T= -C $.
 We can construct two Majorana fields $\chi_1$
  and $\chi_2$ using $\psi$  and  $\psi^c$: $ \chi_1 = (\psi +\psi^c)/\sqrt{2}$,
$ \chi_2 = (\psi - \psi^c)/\sqrt{2}$.
For the Majorana fields the following relation holds
$(\chi_{1,2}) = \pm(\chi_{1,2})^c$.

    It is known that for a particle "a" with charge $q$, spin $S$, and
 mass $m$ the magnetic moment arises with the absolute value
$\mu_a = g({|q|}/{2m})S$, where $g$ is the gyro-magnetic factor or 
the g-factor. The g-factor is usually written as $g =g_0(1+a)$, $a$ is the 
reduced anomalous magnetic moment or anomalous magnetic moment (AMM) of the
particle. For electrons and muons $g_0 = 2$ and DE with AMM have the
so-called Pauli term
\begin{equation}
\quad\qquad\qquad   (\gamma^ip_i^{\prime} - m -
 \frac{iqa}{4m}\sigma^{ij}F_{ij})\psi(x) = 0,
\end{equation}
where $p_i^{\prime} = p_i - qA_i,
\quad F_{ij} = \partial_iA_j  - \partial_jA_i$.

     In the SM framework AMM for electrons
and muons have been evaluated with high accuracy. It is
interesting that the most stringent test of SM and QED is the
comparison of experimental and theoretical values of AMM for
electrons ($a_e$). For experimental determination of $a_e$ the
frequencies of radiation are measured when an electron transits
between quantum levels in a cylindrical Penning trap [5]. The QED
corrections entered into a value $a_e$ are evaluated up to 10th order
(the total contribution of terms of 10th order are only estimated
now) )[6]. If the validity of SM is assumed, the most accurate value
of the fine structure constant  can be obtained at present (0.37
ppb).
\begin{equation}
 g/2(exp) = 1.00115965218073 (28),\quad \alpha^{-1}  = 137.035 999 084 (51)
\end{equation}
The analogous calculations have been made for AMM of a muon [7,
8]  and have been compared with the result of E821 experiment at
Brookhaven National Laboratory (BNL) [9]: 
                   $ a_{\mu}(exp) = 11659208.9(6.3)\times10^{-10}$.
When the SM prediction
                  $ a_{\mu}(SM) = 11659177.3(4.8)\times10^{-10}$
 have been compared to the BNL measurement result  the difference
is obtained
            $ a_{\mu}(exp) - a_{\mu}(SM) = 31.6 (7.9)\times10^{-10}$,
which corresponds to the $4\sigma$ discrepancy! To establish physics beyond
SM  new $g_{\mu} -2$ experiments are planned [10]. The role of
electromagnetic AMMs of quarks should be make clear in the future.

\noindent{\bf 3. The Dirac equation and the relativistic 
quasi-independent quark model}

   The  main  statement  of  the  relativistic  model  for  
quasi-independent  quarks  is that hadron   can  be
described    as  a  system  of  independent
constituents   (or  quasi-independent  ones  with  weak   residual
interactions), which move in some mean self-consistent  field.
 In the framework of RMQIQ a wave function $\psi_i$ for any valent
i-quark  is a solution of DE with the mean field static  potential
$U({\bf r}_i)$.  Energies or mass terms $E_i(n_i^r, j_i)$
 represent the  energies
of constituents in a mean field ($n_i^r$ and $j_i$ are the radial quantum
number   and   the   quantum  number   of   the   angular   moment
correspondingly for the $i-$th constituent) and can be evaluated for
quarks with the help of solutions of the stationary Dirac equation
(for diquarks and constituent gluons with the help of solutions of
the Klein-Gordon-Fock equation):
\begin{equation}
E_i(n_i^r, j_i)\psi_i({\bf r}_i) = [(\alpha_i{\bf p}_i)
+ \beta(m_i+V_0) + V_1]\psi_i({\bf r}_i),
\end{equation}
with  $V_0(r)  = \sigma r/2$ and $V_1(r) = -2\alpha_s/3r$, where
the model  parameters $\sigma$ and $\alpha_s$ have meanings of the
"string tension" and the strong coupling
constant at small distances, correspondingly [3, 4].

     Angular dependence of the single-particle wave functions in a
stationary  state for the spherically symmetric potential  can  be
separated in a well-known manner, namely, the solutions of Eq.(6)
with  the  total angular momentum $j$ and its projection  $m$  can  be
represented as (the subscript $i$ here and below is omitted)
\begin{equation}
\qquad\qquad      \psi({\bf r}) \propto 
\left(\begin{array}{c}f(r)\Omega^m_{jl}({\bf n})\\
-ig(r)(\sigma{\bf n})\Omega^m_{jl}({\bf n})\end{array}  \right),
\end{equation}
where  ${\bf n} =$${\bf r}/r$.
Constituent energies $E_i$ are evaluated by solving the
radial  equations with the model potentials for each  constituent.
If  $k=-\omega (j+1/2)$, where $\omega$ is connected with an eigenvalue
of  the space-parity  operator, the system of the radial  Dirac  equations
 reads
\[
\qquad\qquad                (E - V_0 - V_1 - m)f  =  -\frac{(1-k)}{r}g - g^{'},
\]
\begin{equation}
\qquad\qquad                (E + V_0 - V_1 + m)g  =  -\frac{(1+k)}{r}f + f^{'},
\end{equation}

 One can derive the second order equation  for  the
"large" component $f(r)$ using  Eqs.(8), then making two substitutions
$E = \sqrt{\lambda +m^2}$ and
\[
\qquad\qquad   \phi(r) = rf(r)[V_0(r) - V_1(r) + m + 
\sqrt{\lambda +m^2}]^{-1/2} ,
\]
one comes on to the model radial equation for $\phi(r)$ in the following
form [3, 4]:
\[
  \phi^{"} + \lambda\phi  =  [V_0^2 - V_1^2 + 2(mV_0 +
\sqrt{\lambda +m^2}V_1)  + \frac{k(k+1)}{r^2} +
\]
\[
 \frac{3(V_0^{'}-V_1^{'})^2}{4(\sqrt{\lambda +m^2}-V_1+V_0+m)^2} +
 \frac{k(V_0^{'}-V_1^{'})}{r(\sqrt{\lambda +m^2}-V_1+V_0+m)^2}  -
\]
\begin{equation}
 \frac{3(V_0^{"}-V_1^{"})}{2(\sqrt{\lambda +m^2}-V_1+V_0+m)^2} ]
\end{equation}
This equation is suitable for numerical integration and fitting of
hadron mass spectra. For instance, one can use the
phenomenological mass formula for $q'\bar q$-mesons with the quark and
antiquark energies, which takes into account spin-spin interaction
between quark and aniquark, in the following form [3, 4]:
\begin{equation}
\qquad\qquad  M (q'\bar q) = E_0  + E_1 + E_2 + 
4<\mathbf{s}_1\mathbf{s}_2>V_{12}^{ss}
\end{equation}

     It is interesting to relate the model of quasi-independent
quarks to the constituent quark model, where  the
Zeldovich-Sakharov formula  are frequently used [11, 12]:
\begin{equation}
\qquad\qquad  M (q'\bar q) = m_0  + m_1 + m_2 +  
<\mathbf{s}_1\mathbf{s}_2>v_{12}^{ss},
\end{equation}
where $m_1$, $m_2$ are  masses of constituent quarks, $m_0$ is
some additional phenomenological contribution.  Take into account
the similarity between formulae (10) and (11) we can relate
constituents' energies $E_i$ with constituent quarks' masses $m_i$
within uncertainties of models. Some complications arise for
 exited hadron states. In the model of quasi-independent quarks
$E_i$ remains an energy of $i$-th exited constituent, while in the
constituent quark model an additional contribution representing
energy of excitation must be taken into account.

     The values of parameters $\sigma$ and $\alpha_s$ entered into potentials
$V_0(r)$ and $V_1(r)$ have been determined after the comparison
between experimental and evaluated spectra of the vector and
pseudoscalar mesons. In this manner it was found that the value of $\sigma$
is the same for the light and heavy mesons within the systematic
errors of the model:  $\sigma = (0.20\pm0.01) GeV^2$,
whereas the $\alpha_s$ values run from 0.7 to 0.25 for the light
and heavy mesons, correspondingly [3, 4].  The values
of constituents' energies $E_i$ for $u-, d-, s-$, $c-, b-$quarks
and antiquarks in MeVs  are: $ E_U = 335\pm 2$,
$ E_D = 339\pm  2$, $ E_S = 485\pm  8$, 
$  E_C = 1610\pm 15, E_B = 4952\pm 20$ [13, 14].
So RMQIQ may be considered as the workable generalization
of the constituent quarks model. In the RMQIQ framework the large
number of meson resonance masses have been evaluated, which do not
contradict as the existing experimental data, as the results of
other phenomenological models [4, 15, 16, 17]. In some variants
of the quark model anomalous chromo-magnetic moments of quarks
have been taken into account as well [18, 19, 20].

\noindent{\bf 4. The DGL equation and the Dirac equation in a five-dimensional
Minkowski space}

     It is mentioned above, the spacetime symmetry for quarks can differ from 
the Poincare symmetry. Let us consider some extended kinematical
symmetries and generalized Dirac equations.

     The structure of the algebra of Dirac $\gamma$-matrices (including
the $\gamma_5$ -matrix anticommutating with other $\gamma$-matrices)
shows that the extension of the Poincare group in a 
four-dimensional Minkowski space to a generalized Poincare group in a
five-dimensional Minkowski space is possible. By this way the Dirac
equation is obtained in a five-dimensional Minkowski space [21]:
\begin{equation}
\qquad\qquad\quad         ( \gamma^a p_a  - \kappa)\psi(x) = 0,
\label{dn5}
\end{equation}
where $ p^a    =   i\partial^a$, $ x^a    =  \{t,  \mathbf{x}, x^4\}$,
     $ \gamma^a   =   \{\beta, \beta\alpha, i\gamma_5\}$,
 the metrical tensor $\eta_{ab} = diag \{1, -1, -1, -1, -1\}$.
     FMSDE is invariant under transformations of Poincare group in
a  five-dimensional  Minkowski space $P(1, 4)$,  which  generate
by operators of momentum components $p^a$ and five-dimensional rotations
\begin{equation}
                 J^{'}_{ab} = x_a p_b - x_b p_a  + S^{'}_{ab}, \quad
                 S^{'}_{ab} = \frac{i}{2}\sigma^{'}_{ab},
 \quad \sigma^{'}_{ab}  = \frac{1}{2}(\gamma_a\gamma_b  - \gamma_b\gamma_a).
\end{equation}
\noindent     The  mass of a particle is dependent on a value of an
additional momentum in a new direction $p^4$, 
$m^2=\kappa^2+(p^4)^2$, where $\kappa$ is the constant entered in
Eq. (\ref{dn5}).

     If the symmetry group of a equation was restricted to the
homogeneous pseudoorthogonal  group $O(1,4)$ (or $O(2,3)$), then
the equation is depended on  generators of the pseudoorthogonal
rotations $L_{ab}$  and have the form
\begin{equation}
\qquad         1/2 \gamma^a\gamma^b L_{ab}\psi(x) = \lambda\psi(x)
\end{equation}
 This equation is
proposed by Dirac in 1935 [22]. However $\lambda$ has an imaginary
term due to nonhermitean operators $L_{ij}$ for a finite dimensional
representation. The imaginary term should be single out for a
correct physical interpretation, and the equation (14) was presented
by Gursey and Lee in another form [23]:
\begin{equation}
\qquad         (4i +  \gamma^a\gamma^b L_{ab})\eta(x) = 2mR\eta(x)
\end{equation}
where $\lambda = m R -2i$, $R$ is a radius of a de Sitter space.
Properties of solutions of equations of this type and its application for a
description of elementary particles were studied by Fronsdal, who
used the generalized momentum $p_0 + dp_0L = p_F$, where 
$p_0$ and $L$  have the forms of the usual
generators of translations and  Lorentz transformations in
Minkowski spacetime, and $p_0^2 = m_0^2$, $d = \mu_s/m_0$  [24].  Take
into account the results obtained in Refs. [22, 23, 24], we can
apply the following modified  Dirac-Gursey-Lee equation for a description of
quark characteristics [25]:
\begin{equation}
\qquad   [\gamma_i(p_0^i +dp_0^kL_k^i + i\mu_s\gamma^i/2) + 2i\mu_s S_{ij}
              (L^{ij} + S^{ij})]\psi = m\psi.
\end{equation}
 To estimate a $\mu_s$  value
with the help of a constituent quark mass $m$ and a current quark
mass $m_0$ values,  a quark ground state $\psi_0$ can be used in a meson so
the contribution from $L\psi_0$ can be neglected. In this way we
obtain from Eq.(23) the approximate relation:
 $ m \approx  m_0 + 2i\mu_s$. 
So the constant $\mu_s$  should be pure imaginary and $|\mu|_s\sim 0.16 GeV$
[25].

     The problem of quark and lepton masses is the most important
problem in particle physics now. In the SM framework these masses
arise due to the Yukawa coupling between fundamental fermions and
Higgs field. But, it is known the Higgs meson has not been
discovered [26] and  LHC experiments in the near future must make clear the
question of its existing. The mass problem of quarks and leptons
became more complicated after discovering the neutrino masses [26]. The
neutrino masses are very small, so their origin can differ from the
origin of other fundamental fermion masses due to the fact that
neutrinos can be Majorana particles. It is interesting that values of
quark and neutrino masses can be connected with values of there mixing
 angles [14].

\noindent {\bf 5. The Dirac equation in a quantum phase space}

   It is well known that the Poincare symmetry, which is the 
spacetime symmetry of the  relativistic quantum field theory,
originates from the isotropy and homogeneity of Minkowski 
spacetime and is based on observations of macro- and micro-phenomena
concerning conventional physical bodies and particles. However,
 description of color particles over the whole range
of interaction distances  is possibly needed  using extended symmetries
and theories with new fundamental physical constants (other than
the well known ones $c$ and $\hbar$) in a noncommutative spacetime [27].

 Let us consider the generalized  model for a color particle motion,
when coordinates and momenta are on equal terms and form an eight dimensional
 phase space: $h=$$\{h^{A}|h^{A}=q^{\mu},A=1,2,3,4,$ $ \mu =0,1,2,3,$
$h^{A}=$ $\tau p^{\mu},A=5,6,7,8,$ $ \mu =0,1,2,3\}$,
$P=$ $\{P^{A}|P^{A}=$ $p^{\mu},A=1,2,3,4,$
$ \mu =$ $0,1,2,3,$ $P^{A}=$ $\sigma q^{\mu},A=$ $5,6,7,8,$ $ \mu =0,1,2,3\}$ [28].
The constants $\tau$ and $\sigma$ have dimensions of length and
mass square, correspondingly. Their values can be chosen on the
phenomenological ground or with the help of some functions of
the quantum constants  $\mu$, $\kappa$ and $\lambda$.
We assume  the generalized lenght and mass squares
\begin{equation}
 \qquad\qquad  L^2=h^Ah_A, \quad M^2=P^AP_A
\end{equation}
\noindent are invariant under the O(2,6) transformations,
where $h_A=g_{AB}h^B$,
$g_{AB}=$ $ g^{AB}=$ $diag\{1,-1,-1,-1,1,-1,-1,-1\}$.

Thus the  generalized differential mass squared  can conserve
 for strong interacting color particles:
\[
dM^2 = (dp_0)^2 -(dp_1)^2 - (dp_2)^2 - (dp_3)^2 + \sigma^2(dq_0)^2-
\]
\begin{equation}
\sigma^2(dq_1)^2 -\sigma^2(dq_2)^2 -\sigma^2(dq_3)^2 = (dm)^2+\sigma^2(ds)^2.
\end{equation}
\noindent An important point is that the coordinates $q^{\mu}$ and
the momentum components $p^{\mu}$ are the quantum operators
satisfied the generalized Snyder-Yang algebra (GSYA)
[29, 30, 31, 32]:
\[
[F_{ij}, F_{kl}]=i(g_{jk}F_{il}-g_{ik}F_{jl}+g_{il}F_{jk}-g_{jl}F_{ik}),
\]
\[
[F_{ij}, p_{k}]=i(g_{jk}p_{i} - g_{ik}p_j), 
\quad [F_{ij}, q_k]=i(g_{jk}q_i - g_{ik}q_j),
\]
\begin{equation}
[F_{ij}, I]=0, \quad [p_i, q_j]=i(g_{ij}I + \kappa F_{ij}),
\label{al1}
\end{equation}
\[
[p_i, I]=i(\mu^2q_i - \kappa p_i),   [q_i, I]=i(\kappa q_i-\lambda^2p_i),
\]
\[
[p_i, p_j ]=i\mu^2F_{ij},  \quad [q_i, q_j]=i\lambda^2F_{ij},
\]
\noindent where  $F_{ij}$, $p_i$,
$x_i$ are the generators of the Lorentz group and the operators of momentum
components and coordinates, correspondingly, $I$ is the "identity" operator,
$i, j, k, l = 0, 1, 2, 3$.  The new quantum constants $\mu$
and $\lambda$ have dimensionality of mass and lenght correspondingly.
The constant $\kappa$ is dimensionless in the natural system of units.

By applying the algebra (\ref{al1}) to the description of color particles
the condition $\kappa = 0$ can be imposed. Actually it is known
the  nonzero $\kappa$ leads to the $CP-$violation [31, 32],
but  strong interactions   are invariant
with respect to the $P-$, $C-$ and $T-$transformations on the high level
of precision.  Moreover for color particles one can use the relation
$\mu\lambda = 1$ [28]. In this case  we obtain the
reduction of GSYA to the special Snyder-Yang algebra (SSYA) with
$\mu\lambda = 1$ and $ \kappa = 0$  for strong interaction color particles.
 Denoting $\mu$  as $\mu_c$ and  $\lambda$ as $\lambda_c$ we write
 the following  commutation relations without the standard commutation
relations with Lorentz group generators, which are  shown
 for the GSYA above (see eqs.(\ref{al1})).

\[
\qquad [p_i, q_j]=ig_{ij}I, [p_i, I]=i\mu_c^2q_i,  [q_i, I]=-i\lambda_c^2p_i,
\]
\begin{equation}
\qquad  [q_i, q_j]=i\lambda_c^2F_{ij},  \quad [p_i, p_j ]=i\mu_c^2F_{ij}.
\label{al2}
\end{equation}

We take into account difficulties arised when one try to prove
the confinement on the basis of the QCD first principles [33], and we
simulate  this phemomenon with the help of an
  assumed high symmetry of the nonperturbative OCD interaction beyond the
 Poincare symmetry.
So we turn from the  Poincare symmetry in the
 Minkowski spacetime  to the inhomogeneous O(2,6) symmetry in a
phase space of a color particle [28].

Under these conditions the new Dirac type equation for a spinorial
field $\psi$ has the following form:
\begin{equation}
\qquad  \gamma^AP_A\psi = M\psi, \quad   \gamma^A\gamma^B+\gamma^B\gamma^A = 2g^{AB}.
\label{newd}
\end{equation}
\noindent where
$\gamma^A$ are the Clifford numbers for the spinorial O(2,6) representation.

One can take the product of Eq. (\ref{newd}) with
$\gamma^AP_A+M$ and apply Eqs. (\ref{al1}), then the following
equation for $\psi$ can be obtained
\[
\qquad (p^ip_i + \sigma^2q^iq_i + 2\Sigma_{i<j}S^{ij}F_{ij} +
\]
\begin{equation}
\qquad  + 2\sigma S^0I)\psi = M^2\psi,  S^{0}=\frac{i}{2}C^{0},
S^{ij}=\frac{i}{2}C^{ij},
\label{newds}
\end{equation}
\noindent where $C^0$, $C^{ij}$ are the combinations of the
$\gamma$-matrices and the constants $\lambda$, $\mu$ and $\kappa$
[28]. Eq. (\ref{newds}) contains the oscillator potential, which restricts
a motion of a color quark. Besides that we broke the inhomogeneous
 O(2,6) symmetry with the help of the commutation relations (\ref{al1}).
In the special case $\mu\lambda = 1$, $ \kappa = 0$ and the commutation
 relations (\ref{al2}) for SSYA we will obtain more
simple expressions for the $C^0$ and $C^{ij}$, but the form of the
Eq. (\ref{newds}) will remain unchanged. Note that  Eq. (\ref{newds}) can also be
applied for a description of a confinement of boson particles such as
diquarks and gluons with the same confinement parameter $\sigma$.

One can get an  estimation of the $\sigma$ value
  using Eq. (\ref{newds}). As it is seen,
$M^2$ and $p^2$ entered into Eq.(\ref{newds})
can be considered as current and constituent  quark masses squared,
respectively.  So Eq. (\ref{newds}) indicates that the convential
relation $p_{cur}^2=M^2$ for a current quark should be transform
to $p^2=M^2+ \Delta^2$ for a constituent quark, where $m^2=M^2+\Delta^2$
is a constituent mass squared.
To estimate the $\sigma$ value with the help of
a constituent quark mass $m$ and a current quark mass $M$  values
 a ground state $\psi_0$  in a meson has been considered  neglecting
 the orbital angular momentum contribution $L\psi_0$, as it had been done for
DGLE. We can also estimate the $\sigma$ value
with the help of the value of the  confinement rising  potential coefficient
[3].
It is seen from Eq. (\ref{newds}) that the coefficient of the oscillator
confinement  potential for a color particle is equal to $\sigma^2$.
Within the potential approach this coefficient is connected
with the string tension  $\sigma_{str}$  typically as $\sigma^2=\sigma_{str}^2/4$,
where $\sigma_{str}$ varies from 0.19 $GeV^2$ to 0.21 $GeV^2$ [3].
 Hence $\sqrt{\sigma}$ $\approx 0.3 GeV $  and
$\lambda_{conf}\approx 0.6 Fm$. Clearly  it is assumed that the
conceptions of the asymptotical  oscillator potential and the
constituent quark are applicable in a confinement domain.

\smallskip
\noindent{\bf 6. Conclusion}
\smallskip

The Dirac equation is successfully employed for the description of
particles with the spin $1/2$. If one take into account  an additional
Pauli term, then  the Dirac equation describes in the SM framework
the modification of a magnetic moments of leptons due to radiative
corrections. However, someone has cast doubt on the relevance of
DE for quarks in a confinement domain, i.e. in  the nonperturbative domain of
QCD.  It is very likely that in this case we should  modify
essentially the Dirac equation or to pass on to a new equation of
the Dirac type.
     The possible generalizations  of the Dirac equation have been
considered in this report. The Dirac equation in a quantum phase
space is the most interesting from the standpoint of  description
of quark motion. This equation involves an oscillatory potential, so it
can play a role of a model of  color particle confinement. 

      The  author  is grateful to \fbox{Yu.V.  Gaponov}, Ya.I. Azimov, S.V. Semenov
and A.M. Snigirev  for valuable discussions and the organizers of
the XLIV PNPI Winter School for the hospitality.

\newpage

\noindent{\large\bf References}

\smallskip

  1.  N.N. Bogoliubov, A.A. Logunov, A.I. Oksak, and I.T. Todorov,
     {\it General principles of quantum field theory.} Moscow, Nauka, 1987
     (in Russian).

 2. F.J. Yndurain, 
{\it Quantum Chromodynamics.} Springer-Verlag, New York-Berlin-
Heidelberg-Tokyo, 1983.

3.  V.V. Khruschov, V.I. Savrin, and S.V. Semenov,  Phys. Lett.
B525, 283 (2002);
  V.V. Khruschov, and S.V. Semenov,  Part.Nucl.Lett. 5, 5 (2002).

4.  V.V. Khruschov, 
Proc. XLI PNPI Winter School, P. 139,  S.- Peterburg, 2008.  

5.  D.Hanneke, S.Fogwell, and G.Gabrielse,  Phys.Rev.Lett.
100, 120801(2008).

6.  T. Aoyama et al., Phys. Rev. Lett. 99, 110406 (2007).

7.  A.L. Kataev, Phys. Rev. D74, 073011 (2006).

8. T. Teubner et al., Chin. Phys. C 33, 11 (2009),
arXiv:1001.5401.

9. G.W. Bennett et al., Phys. Rev. D73, 072003 (2006).

10. D.W. Hertzog, Nucl. Phys. B (Proc. Suppl.), 181-182 (2008).


11.  Ya.B. Zeldovich, A.D. Sakharov, Yad. Fiz. 4, 395 (1966).

12.  A.D. Sakharov, ZHETF 78, 2112 (1980).

13. M.D.  Scadron, R. Delbourgo, G. Rupp, J. Phys. G 32, 735
(2006).

14.  Yu.V.Gaponov, V.V.Khruschov, S.V.Semenov, Yad.Fiz. 71,
163(2008).

15.  De-Min Li and Shan Zhou, Phys. Rev. D78, 054013 (2008).

16.  A. Duviryak, SIGMA  4, 048 (2008).

17.  R.R. Silbar, T. Goldman, arXiv:1001.2514.

18.  H.J. Schnitzer, Phys. Rev. D19, 1566 (1979).

19.  B.A. Arbuzov et al., Yad. Fiz. 42,  987 (1985).

20.  V.O.Galkin, A.Yu.Mishurov, R.N.Faustov, Yad.Fiz. 55, 2175(1992).

21.  W.I. Fushchich, I.Yu. Krivsky, Nucl. Phys. B 14, 321 (1969).

22.  P.A.M. Dirac,  Ann. Math. 36, 657 (1935).

23.  F. Gursey, T.D. Lee, Proc. Nat. Acad, Sci. (USA), 49, 179
(1963).

24.  C.Fronsdal,  Rev.Mod.Phys. 37, 221(1965); Phys.Rev. D10,
589(1974).

25.  V.V. Khruschov,  Proc. XIII Inter. Conf.  Select. Prob.
Mod. Phys. dedic. 100th anniv. of the birth of D.I.Blokhintsev, p. 224,
JINR, 2009.

26. C. Amsler et al., Phys. Lett. B667, 1 (2008).

27.  M. Toller, Phys. Rev. D70,  024006 (2004).

28.  V.V. Khruschov, Grav. Cosmol. 15, 323 (2009);
arXiv:0912.2819.

29.  H. Snyder, Phys. Rev. 71, 38 (1947).

30.  C.N. Yang,  Phys. Rev. 72, 874 (1947).

31.  A.N. Leznov and V.V. Khruschov, Prep.IHEP 73-38, Serpukhov (1973).

32.  V.V. Khruschov, A.N. Leznov, Grav. Cosmol. 9, 159 (2003).

33.  J. Greensite, Prog. Part. Nucl. Phys. 51, 1 (2003).

\end{document}